\def\spose#1{\hbox to 0pt{#1\hss}} 
\def\gta{\scriptsize\mathrel{\spose{\lower
2pt\hbox{$\mathchar"218$}}
     \raise 3.0pt\hbox{$\mathchar"13E$}}\normalsize}

\documentstyle[12pt,rotate,portland,psfig,epsf,aaspp4]{article}
\input{rotate}
\lefthead{Guo \& Burrows}
\righthead{ASCA Observations of SNR VRO42.05.01}
\received{}
\accepted{13 February 1997}
\journalid{}{}
\articleid{}{}

\newcommand{\asca}{\mbox{\it ASCA}}
\newcommand{\degrees}{\mbox{$^{\circ}$}}    

\newcommand{\kms}{\mbox{\rm km s$^{-1}$}}
\newcommand{\NH}{\mbox{${\it N}_{\rm H}$} } 

\newcommand{\rosat}{{\it ROSAT}}
\newcommand{\tq}{\mbox{$\frac{3}{4}$}}	    
\normalsize

\begin{document}
\title{ASCA Observations of the Supernova Remnant VRO42.05.01}
\author{Zhiyu Guo and David N. Burrows}
\affil{Department of Astronomy \& Astrophysics, Pennsylvania State University,\\
525 Davey Lab, University Park, PA 16802; \\
I: guo@astro.psu.edu \\
I: burrows@astro.psu.edu}

\begin{abstract}
We present the results of our {\asca} SIS and GIS observations of
the supernova remnant VRO42.05.01.  
Our spectral fits indicate that the SNR is isothermal 
($\sim 8.3 \times 10^{6}$~K), consistent with our {\rosat} 
analysis results.  The absorbing column density 
($\sim 2.9 \times 10^{21}$~cm$^{-2}$) obtained
from these spectral fits is much smaller than expected for the nominal distance 
of 5 kpc,
indicating that the line of sight toward VRO42.05.01 has an unusually low gas 
density.  
The spectral resolution of {\asca} allows us to determine elemental
abundances for the hot X-ray emitting plasma in the bright ``wing''
component of this remnant.
We find that Mg, Si, and Fe are
underabundant, and attribute these low abundances to the galactic metallicity
gradient and to the location of the remnant in the outer Milky Way.
\end{abstract}


\section{Introduction}
VRO42.05.01 (G166.0+4.3) is classified as a composite supernova remnant 
(SNR), with a shell-like radio morphology and a centrally-peaked
X-ray morphology.  
Radio continuum observations (Landecker et al. \markcite{1} 1982; 
Pineault, Landecker, \& Routledge \markcite{2} 1987)
reveal an unusual shape, with a northeastern circular 
component (the ``shell'') intersected by a much larger bowl-shaped component 
in the southwest (the ``wing'' component), 
which bears
a remarkable resemblance to numerical models by Tenorio-Tagle,
Bodenheimer, \& Yorke \markcite{3} (1985) of blast waves expanding
across a density discontinuity.
The morphology suggests that this
SNR has broken out of the cloud within which it formed, expanded across an
interstellar tunnel or cavity, and is now interacting with the material that
forms the opposite tunnel wall (Pineault et al. \markcite{2} 1987).  

Optical observations find line ratios in the optical filaments of VRO42.05.01
characteristic of supernova remnants (Fesen, Blair, \& Kirshner \markcite{4} 1985; Pineault et al. \markcite{5} 1985)
and a filamentary structure similar to that observed in the radio
continuum (Fesen, Gull, \& Ketelsen \markcite{6} 1983; Pineault et al. \markcite{5} 1985).  Shock velocities of 
over 100 {\kms} are implied by the optical data 
(Fesen et al. \markcite{6} 1983; Lozinskaya \markcite{7} 1979). 

Burrows \& Guo \markcite{8} (1994, hereafter Paper I) presented the first X-ray images and spectra 
of VRO42.05.01, based on {\rosat} PSPC observations.  We found that the overall
shape of the X-ray remnant is similar to that of the radio remnant, but 
that the X-ray image differs
from the radio morphology in that the former is centrally-peaked instead
of edge-brightened.
The X-ray morphology is consistent with the evaporating cloud model of
White \& Long \markcite{10} (1991).
We also found that the X-ray data are consistent with an isothermal remnant
($\sim 8.5 \times 10^{6}$ K) with variations in absorbing column density
($\sim$ a few $\times 10^{21}{\rm cm^{-2}}$) across the remnant.  We noted
that the absorbing column obtained from the X-ray data is surprisingly small
for the nominal distance of 5 kpc, which requires a low mean density along this
line of sight.

In this paper, we present the results from our analysis of data 
obtained from two Solid-state Imaging Spectrometer (SIS) detectors and
two Gas Imaging Spectrometer (GIS) detectors on board the Japanese X-ray
satellite {\asca} (Tanaka, Inoue, \& Holt \markcite{9} 1994).   We shall describe the 
data acquisition 
and analysis in
\S 2, discuss the physical implication of the data analysis results and a
comparison with former results in \S 3,
and present our conclusions in \S 4.

\section{Data and Analysis}

VRO42.05.01 was observed by the {\asca} between 9 March 1994 and 
11 March 1994 with 
two SIS detectors (4 CCD mode) and two GIS detectors (PH mode) in use 
simultaneously.  Three pointings
covered virtually the entire remnant with 
exposure times of 20 ks to 30 ks in each pointing.
Standard event screening has been applied to both SIS and GIS observations
using XSELECT (V1.2).  The screening procedures include removal of hot and 
flickering pixels from the SIS observations, selection of usable field and
background rejection based on rise time interval for the GIS observations,
and selection of good observation time intervals.  
Hot and 
flickering pixel counts are found to comprise more than 95\% of the total 
counts from each CCD chip, and more than 99\% in some CCD chips.  Cleaned SIS and GIS images of the SNR are shown in 
Figures~\ref{fig:SIS image} and \ref{fig:GIS image} (Plates 00 and 00).

Even in the
CCD chip that covers the brightest portion of the SNR, there are less
than 3500 counts left after cleaning the data.  
This has made our original goal of studying
the spectral properties across the SNR on angular scales smaller than the
detector size impractical.  
Instead, we have combined all of the events from each CCD to produce
spectra of $11' \times 11'$ sized regions of the remnant.
Only three pairs 
of the SIS CCD chips that cover the brightest portions of the SNR
were able to provide spectra with sufficient signal to noise to
allow spectral fits with reasonable statistics.  These three CCD chip pairs are
SIS0~CCD2~/~SIS1~CCD0 in the first pointing
(marked as Region 1 in Figure~\ref{fig:SIS image}), SIS0~CCD0~/~SIS1~CCD2 
in the third pointing (Region 2), and SIS0~CCD3~/~SIS1~CCD1 
in the second pointing (Region 3).  
The GIS 
spectra were extracted from these same three regions.  
The energy ranges of both SIS and GIS spectra are from 0.6 keV to
$\sim$ 2.2 keV (gold edge).  Blank sky event lists created by the {\asca} GOF
were used for background subtraction in all the spectral fits due to 
the limited number of
background counts in our observations.  Spectra from the GIS3 were 
binned to 128 PI channels (instead of 256) because
of the telemetry problem for the GIS3 data in this particular observation
period.
No distinct emission lines are seen in any of the spectra.  We show the SIS0
spectrum from Region~1, which covers the brightest portion of the SNR, in
Figure~\ref{fig:spec_model}.

Spectral fits to a variety of thermal models were performed 
using XSPEC (V9.01) for both the SIS and the GIS data sets.  
We were unable to obtain acceptable results in simultaneous fits to 
data from the same region of the SNR from all
four detectors, suggesting an inconsistency between
the SIS and the GIS.  We therefore only used the two SIS spectra
from each pointing and 
obtained reasonable results after adjusting the detector gain by a few percent. 
We tried to fit our spectral data using single temperature
and two temperature Raymond-Smith models with solar
abundances but were not able to obtain acceptable fits ($\chi^{2}$/d.o.f. are
299/112 and 282/110, respectively).
The centrally-peaked X-ray morphology and lack of emission
lines are suggestive of a nonthermal source, but we could not obtain an 
acceptable fit to a power law spectrum ($\chi^{2}$/d.o.f. is 314/112).
We then tried using a non-equilibrium ionization (NEI) model with solar 
abundances produced by Richard Edgar at CfA, but still could not obtain
a satisfactory fit ($\chi^{2}$/d.o.f. is 251/111).
The SIS spectra and images were compared with a 2-D
hydrodynamical model produced by 
Jon Slavin at GSFC, which modeled a blast wave crossing a density
discontinuity and integrated the plasma emissivity along various lines of
sight to produce spectral simulations.  With appropriate model parameters
(explosion energy and location, and ISM density contrast) this model
could approximate either the spatial appearance of VRO42.05.01 or its 
approximate spectral
shape, but we were unable to satisfy both spatial and spectral observations
simultaneously. 

We were finally able to obtain acceptable spectral fits using a 
Raymond-Smith model with variable abundances.  The
fit results of the SIS spectra from the three brightest portions of the SNR
are given in Table 1 and are shown in Figures~\ref{fig:spec_model}--~\ref{fig:abund_contour} .  
The best fit model spectrum for Region~1 is plotted as a heavy histogram
overlaid on the data in 
Figure~\ref{fig:spec_model}.
Low abundances of Mg, Fe, and Si are required to reproduce the nearly
featureless observed spectrum.  By comparison, the light histogram
in Figure~\ref{fig:spec_model} shows the best fit Raymond-Smith model
with solar abundances, which is too faint below 0.8 keV and has discrepancies
at 1.26 keV and 1.85 keV associated with lines from Mg and Si which are 
prominent in the model but are not 
observed in the data.
Confidence contour plots of the abundances 
(Figure~\ref{fig:abund_contour}) show that all
three elements must be below solar abundance in the wing component
for the Raymond-Smith model to successfully fit the data.

\section{Discussion}

The {\asca} spectral fit results presented in
Table~\ref{tab:fit results} and Figure~\ref{fig:nh_T} confirm 
our previous ROSAT results that the SNR is isothermal and
that the absorbing column densities towards Region 1
and Region 2 are consistent with each other but quite different from
that towards Region 3.
%
%
%
%
The temperatures obtained here are in reasonable agreement 
with those determined on the basis of our {\rosat} observation.  
The emission measures obtained here cannot be directly compared with
those from Paper I, because the regions included in the fit are not
identical.
The column density results are puzzling.  Both data sets indicate a 
substantial difference in the absorbing column between the shell and wing
components, but the
sense is reversed: the {\rosat} data indicate a smaller {\NH} for the wing
component, while the {\asca} data indicate a smaller {\NH} for the shell
component.  We are inclined to trust the {\rosat} data in this case because
the fitted column density is sensitive to the calibration of the lowest {\asca}
channels, which may not be entirely secure.  However, it is possible that the 
absorbing gas is patchy and that the discrepancy found here is real.  In 
any case, these column densities imply a low mean density for this
line of sight, for the reasons discussed in Paper I.  Further observations may 
be required to resolve this discrepancy.

Unlike {\NH} confidence contours
(Figure~\ref{fig:nh_T}), contours of elemental abundances
(Figure~\ref{fig:abund_contour}) do not show clear differences between
the wing and shell components.  
The confidence contours indicate that 
Mg and Si are underabundant in
the wing component (Regions 1 and 2), while the abundances of these 
elements in the
shell component are more uncertain and could be consistent with either solar
abundances or with the depleted abundances found in the wing. 
The Fe abundances appear to be below solar values in both the
wing and shell components.
As shown in Figure~\ref{fig:spec_model}, the low abundances are a 
consequence of the nearly featureless spectrum of this remnant.
This is unusual, as most thermal SNRs have strong lines in their
X-ray spectra.  

VRO42.05.01 is an old SNR (2.4 $\times$ $10^{4}$ yr, Paper I) and its X-ray 
emission should be dominated by emission from the 
swept-up ISM.  The low abundances we find are therefore suggestive of low 
abundances in the local interstellar environment of the SNR.
Since VRO42.05.01 is estimated to be 5 kpc from the Sun at galactic longitude 
166{\degrees}, the radial gradient of elemental abundances in the Milk Way
may be responsible for the low abundances of this remnant.
For a galactic radial metallicity gradient of -0.05 dex ${\rm kpc^{-1}}$
(Pagel \markcite{11} 1981) in the neighborhood of the Sun, 
we estimate that the metallicity at the location of the 
remnant is 56\% of solar.  This is within the confidence contour ranges
for Fe and Mg, and within a factor of 2 of the allowed range for Si.  
Taking into account the complication that galactic arms 
have higher elemental gradients (as large as -0.2 dex ${\rm kpc^{-1}}$ 
[Talent \& Dufour \markcite{12} 1979]), we would expect even 
lower abundances that are closer to our best fit values.

\section{Conclusions}

We conclude that the analysis of the {\asca} data has shown consistent results
with those from our {\rosat} analysis.  The SNR VRO42.05.01 is found to be
isothermal
with variations of neutral hydrogen column density across the remnant.  The
low column density suggests that the line of sight toward the SNR has an
unusually low gas density.  The important new result from the {\asca} data is
the first measurement of abundances in the hot plasma.
The elemental abundances of Mg, Si, and Fe
are found to be less than the solar abundance in the wing component 
where we have good statistics, but have a
large uncertainty in the shell component where we have fewer counts.  
We attribute the low abundances to the galactic metallicity gradient and the
resultant low metallicity in the swept-up ISM.  This may explain the spectral
differences between this remnant and 3C400.2, which has many similarities
to VRO42.05.01 but has strong emission lines of Mg, Si, and Fe in its {\asca}
spectrum (Guo et al. \markcite{15} 1997).

\acknowledgements
We would like to thank Eric Gotthelf at {\asca} GOF for useful discussions on
making the images, and Richard Edgar and Jon Slavin for providing NEI
spectral models for comparison with our data.
This work has been sponsored by NASA grant NAG5-2537.

\clearpage


\figcaption[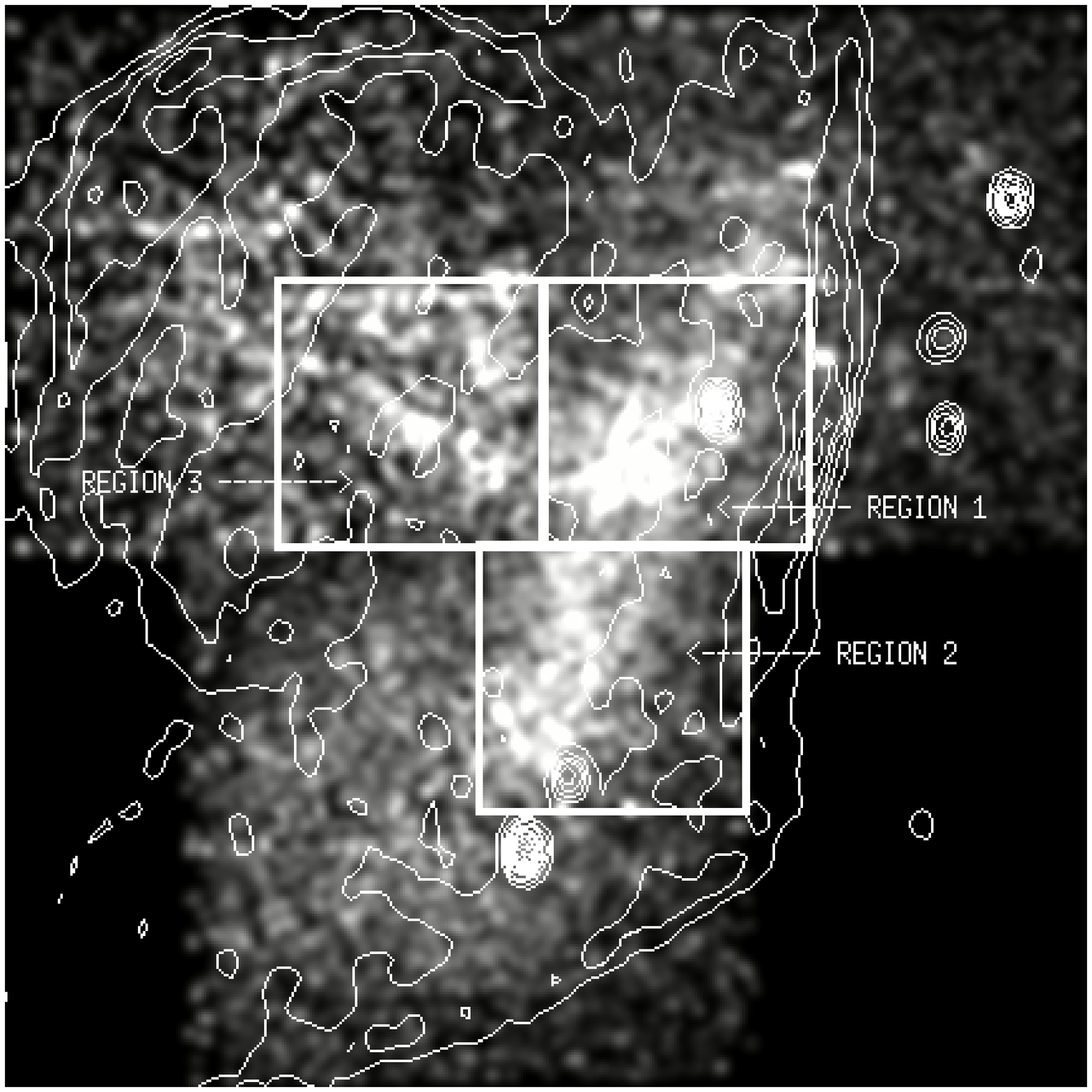]{
The ASCA SIS image of VRO42.05.01 overlaid with 1420 MHz radio
continuum contours (courtesy T. Landecker).  
The X-ray mosaic, which was smoothed 
with a $3'$ (FWHM) Gaussian, includes three
pointings and combines counts from both SIS0 and SIS1.  
The three regions from which we extracted
spectra are marked as boxes and are labeled Regions 1, 2, and 3.
        \label{fig:SIS image}}

\figcaption[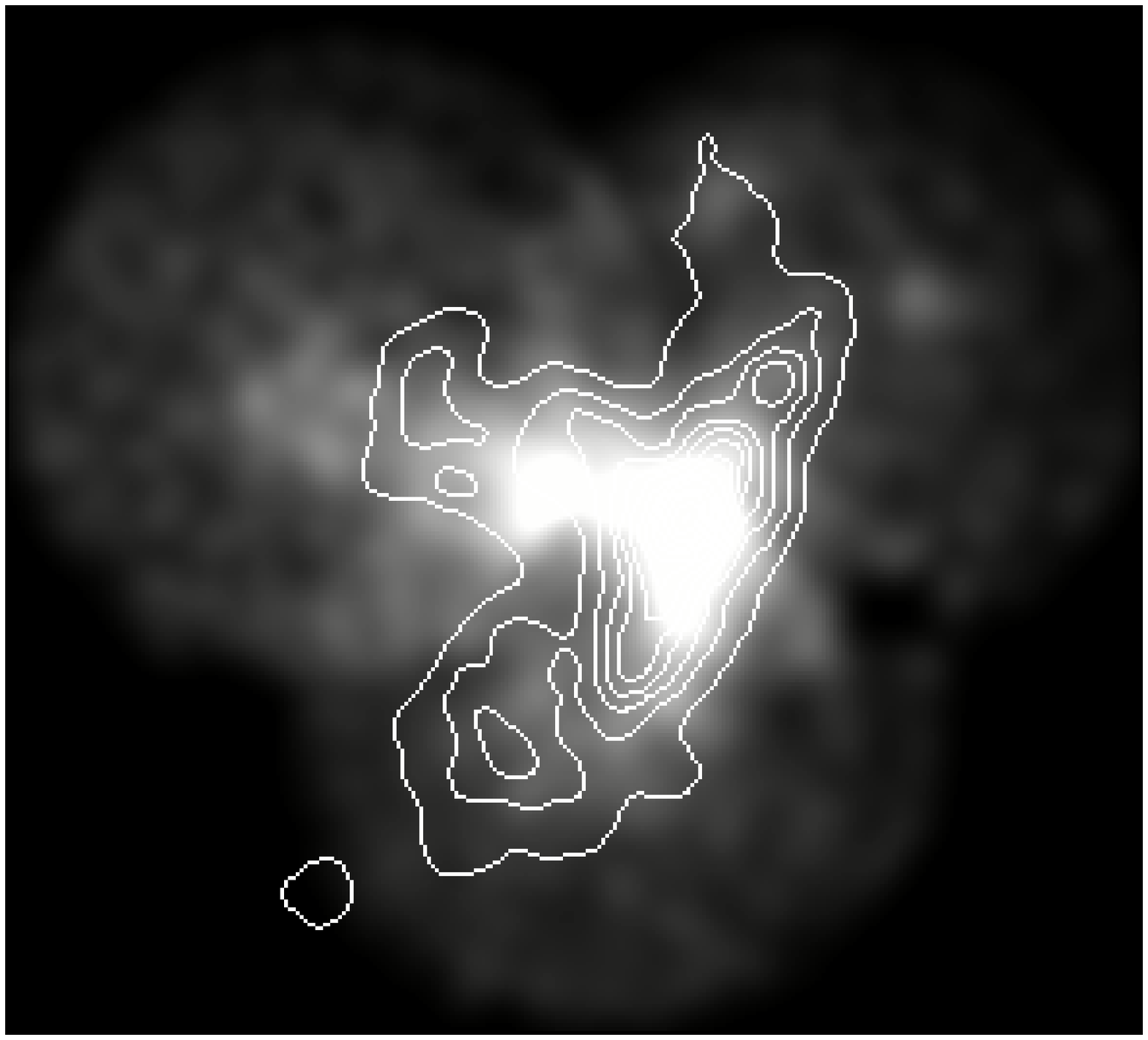]{
Mosaic of {\asca} GIS images from three pointings, smoothed with a
$4.25'$ (FWHM) Gaussian. Overlaid contours are from the {\rosat} {\tq} keV
image (Paper I).
        \label{fig:GIS image}}

\figcaption[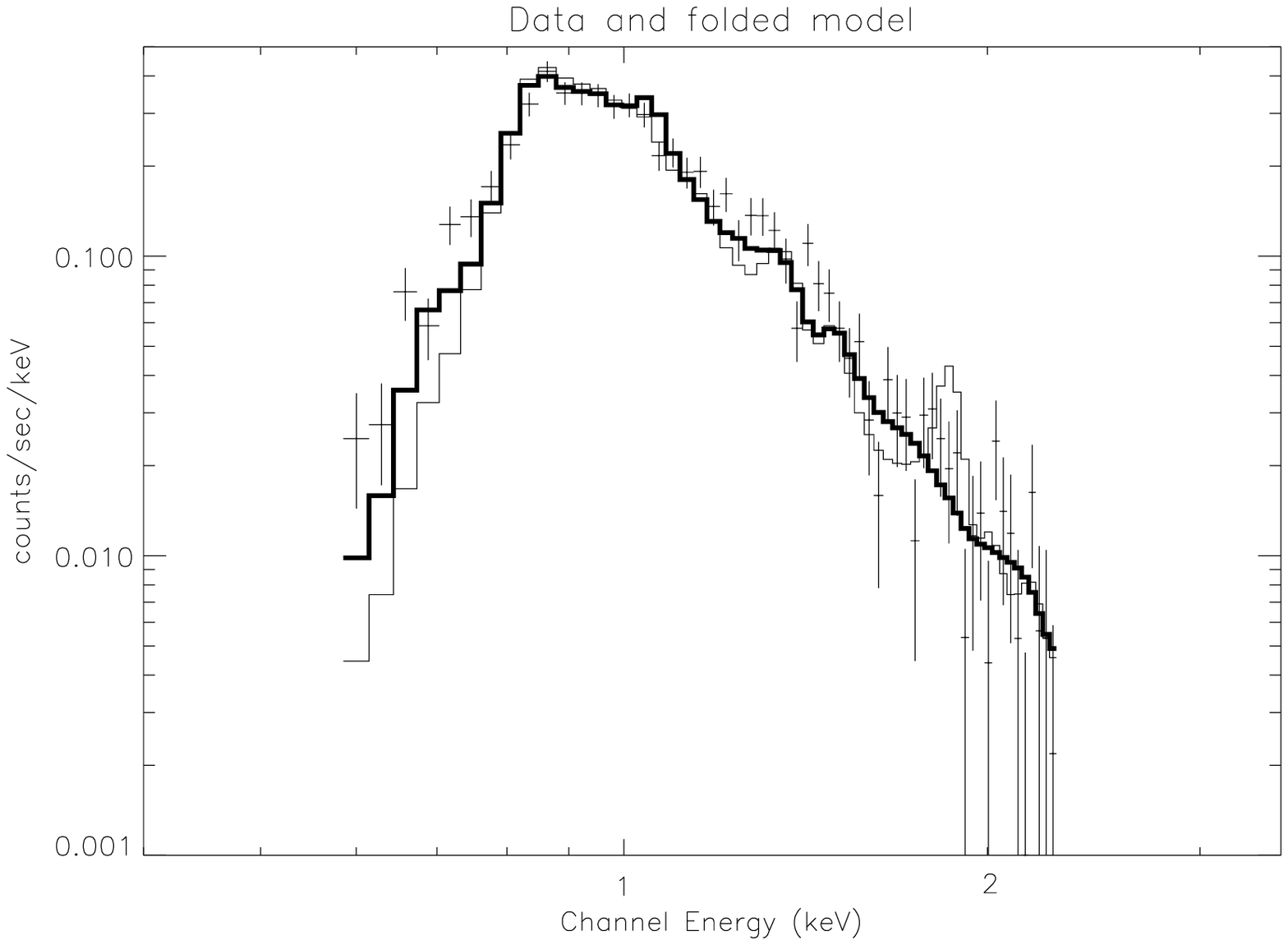]{
SIS spectrum extracted from CCD chip 2 on SIS0 (Region 1).
The data points are overlaid with the best
Raymond-Smith model fit using a single temperature plasma with
variable abundances (thick line) (Table 1), and the best Raymond-Smith model
fit using a single temperature plasma with solar abundances (thin line).
	\label{fig:spec_model}}


\figcaption[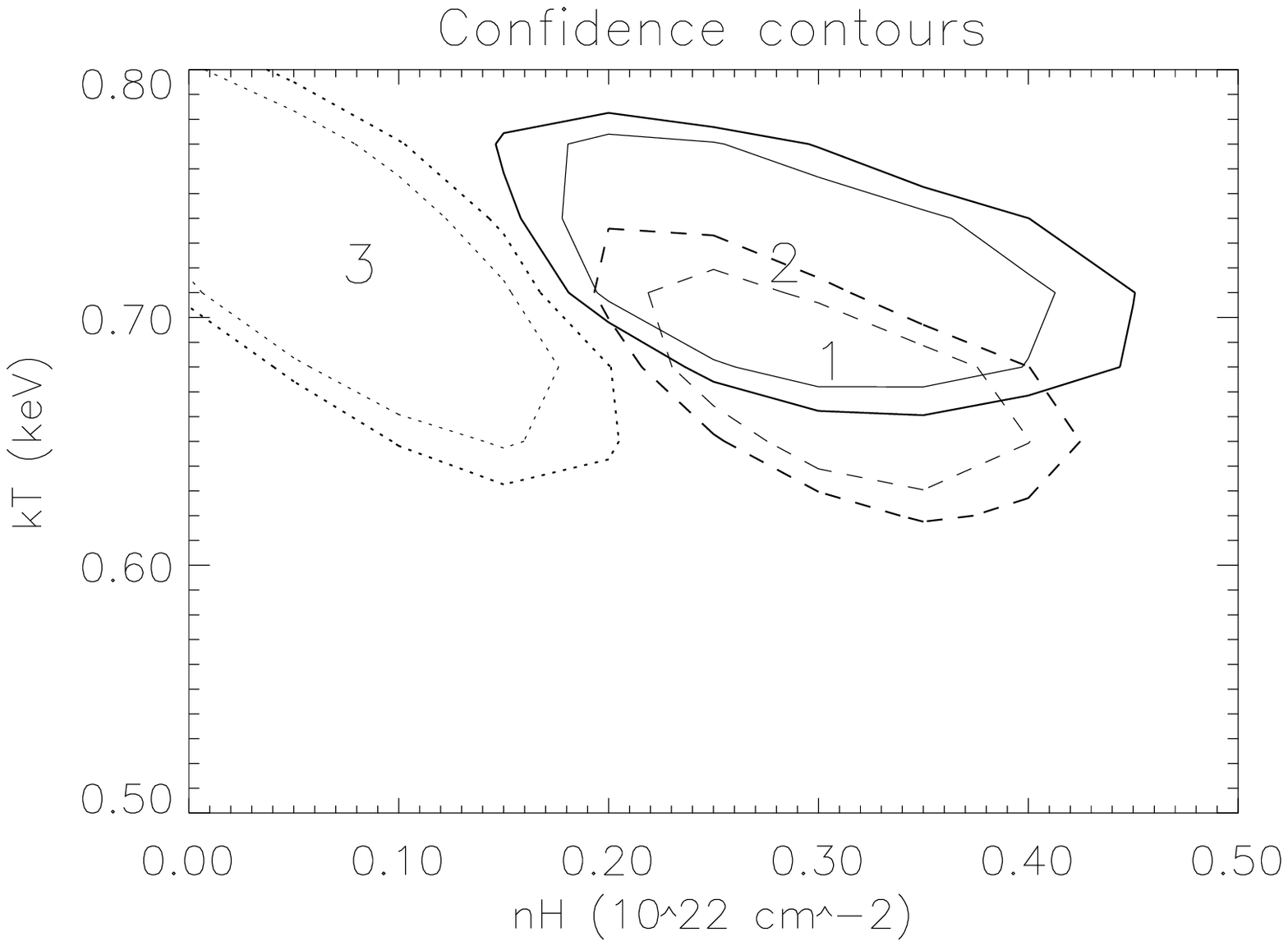]{
Confidence contours for the fit parameters temperature (T) and
absorbing column (nH).  Dashed, solid, and dotted contours represent
Regions 1, 2, and 3, respectively.  Numbers `1', `2', and `3' mark
the best-fit values in each contour for the corresponding regions.  The
inner contours represent a confidence level of 70\% ($\chi^{2}_{min}$ + 7.23)
while the outer contours
represent a confidence level of 90\% ($\chi^{2}_{min}$ + 10.64).  
	\label{fig:nh_T}}

\figcaption[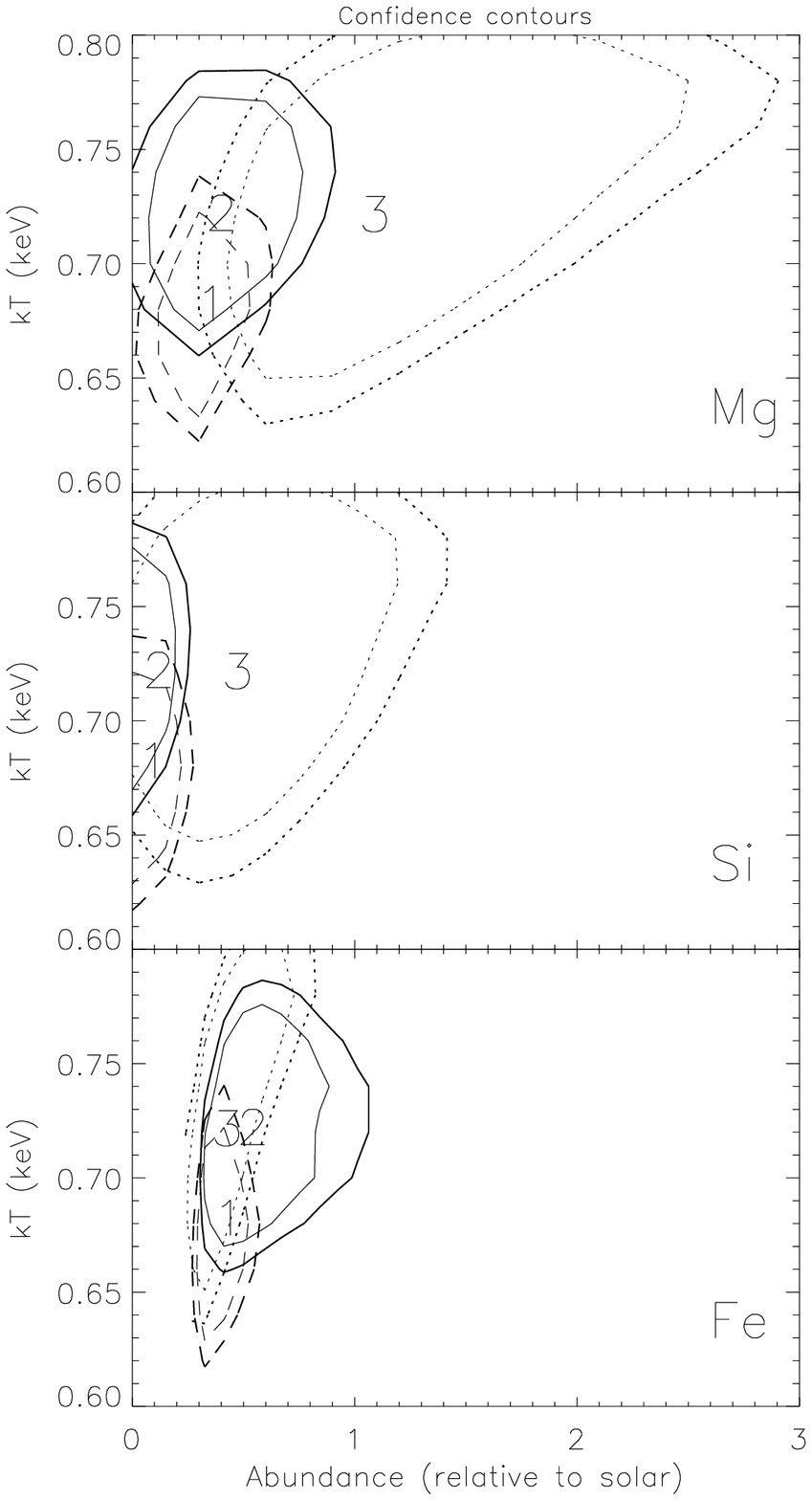]{
Same as Figure 4, but for elemental abundances of Mg (upper), Si (middle),
and Fe (lower) (relative to the solar abundance).
	\label{fig:abund_contour}}





\clearpage

\begin{deluxetable}{rcccccc}
\tablecolumns{7}
\tablewidth{0pt}
\small
\tablecaption{X-ray spectral fit results from the {\asca} SIS and the {\rosat} PSPC data
\label{tab:fit results}}
\tablehead{
\colhead{} & \multicolumn{3}{c}{{\asca} SIS} & \colhead{} 
	& \multicolumn{2}{c}{{\rosat} PSPC\tablenotemark{b}} \\ 
\cline{2-4} \cline{6-7} \\
\colhead{Parameters} & \colhead{Region 1\tablenotemark{a}} & \colhead{Region 2\tablenotemark{a}} & \colhead{Region 3\tablenotemark{a}} & \colhead{} & \colhead{} & \colhead{} \\
\colhead{CCD Pairs used} & \colhead{S0C2 / S1C0} & \colhead{S0C0 / S1C2} & \colhead{S0C3 / S1C1} & \colhead{} & \colhead{} & \colhead{} \\
\colhead{} & \colhead{Wing} & \colhead{Wing} & \colhead{Shell} & \colhead{} & \colhead{Wing} & \colhead{Shell} 
}
\startdata
Temperature ($10^{6}$ K)            & 7.8\phn\phn & 8.3\phn & 8.3\phn & & 8.1 & 8.6 \nl
{\NH} ($10^{21}$ cm$^{-2}$)         & 2.9\phn\phn & 2.8\phn & 0.7\phn & & 1.3 & 2.6 \nl
Normalization ($10^{11}$ cm$^{-5}$) & 4.6\phn\phn & 3.1\phn & 1.6\phn & & 4.7 & 3.0 \nl
Mg abundance                        & 0.27\phn    & 0.33    & 1.02    & & N/A & N/A \nl
Si abundance                        & 0.026       & 0.0\phn & 0.40    & & N/A & N/A \nl
Fe abundance                        & 0.37\phn    & 0.47    & 0.36    & & N/A & N/A \nl
$\chi^{2}$/d.o.f.		    & 194/109     & 167/96  & 158/97  & & 12.6/15 & 3.75/15 \nl 	
\enddata
\tablenotetext{a}{The three spectral regions are marked in Figure~\ref{fig:SIS image}.}
\tablenotetext{b}{The parameters for the ROSAT data are from Burrows and Guo (1994).}
\end{deluxetable}

\clearpage


\begin{figure}
\epsscale{0.8}
\plotone{asca_sis+radio_contour.ps}
{Fig.~\ref{fig:SIS image}}
\end{figure}

\clearpage

\begin{figure}
\epsscale{0.8}
\plotone{vro_gis_mo_rosat_contour_bw.ps}
{Fig.~\ref{fig:GIS image}}
\end{figure}
\clearpage




\begin{figure}
\epsscale{1.0}
\plotone{spec_model.ps}
{Fig.~\ref{fig:spec_model}}
\end{figure}

\clearpage

\begin{figure}
\epsscale{1.0}
\plotone{SIS_nh_T_final.ps}
{Fig.~\ref{fig:nh_T}}
\end{figure}

\clearpage

\begin{figure}
\epsscale{0.6}
\plotone{abund_contour.ps}
{Fig.~\ref{fig:abund_contour}}
\end{figure}

\clearpage








\end{document}